\documentclass[sigconf]{acmart}
\usepackage{fdiaz}

\usepackage{booktabs} % For formal tables
\setcopyright{rightsretained}
\usepackage{makecell}
\usepackage{tcolorbox}
\usepackage{subcaption}
\usepackage{tablefootnote}
\usepackage{arydshln}
\usepackage{colortbl}
\usepackage{xcolor}
\usepackage{graphicx}
\usepackage{epstopdf}

\definecolor{lightgray}{gray}{0.9}
\definecolor{Gray}{gray}{0.9}
\definecolor{highlight}{rgb}{0.9, 0.9, 0.9} % Light gray

\AtBeginDocument{%
  }

\copyrightyear{2025}
\acmYear{2025}
\setcopyright{cc}
\setcctype{by}
\acmConference[SIGIR '25]{Proceedings of the 48th International ACM SIGIR Conference on Research and Development in Information Retrieval}{July 13--18, 2025}{Padua, Italy}
\acmBooktitle{Proceedings of the 48th International ACM SIGIR Conference on Research and Development in Information Retrieval (SIGIR '25), July 13--18, 2025, Padua, Italy}
\acmDOI{10.1145/3726302.3730335}
\acmISBN{979-8-4007-1592-1/2025/07}

\makeatletter
\gdef\@copyrightpermission{
  \begin{minipage}{0.3\columnwidth}
   \href{https://creativecommons.org/licenses/by/4.0/}{\includegraphics[width=0.90\textwidth]{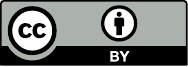}}
  \end{minipage}\hfill
  \begin{minipage}{0.7\columnwidth}
   \href{https://creativecommons.org/licenses/by/4.0/}{This work is licensed under a Creative Commons Attribution International 4.0 License.}
  \end{minipage}
  \vspace{5pt}
}
\makeatother



\begin{document}

\title{Tip of the Tongue Query Elicitation for Simulated Evaluation}

\author{Yifan He}
\authornotemark[1]
\affiliation{%
  \institution{Carnegie Mellon University}
  \city{Pittsburgh}
  \state{PA}
  \country{USA}
}
\email{yifanhe@alumni.cmu.edu}

\author{To Eun Kim}
\authornote{Equal Contribution.}
\affiliation{%
  \institution{Carnegie Mellon University}
  \city{Pittsburgh}
  \state{PA}
  \country{USA}
}
\email{toeunk@cs.cmu.edu}

\author{Fernando Diaz}
\affiliation{%
  \institution{Carnegie Mellon University}
  \city{Pittsburgh}
  \state{PA}
  \country{USA}
}
\email{diazf@acm.org}

\author{Jaime Arguello}
\affiliation{%
  \institution{UNC Chapel Hill}
  \city{Chapel Hill}
  \state{NC}
  \country{USA}
}
\email{jarguell@email.unc.edu}

\author{Bhaskar Mitra}
\affiliation{%
  \institution{Microsoft Research}
  \city{Montréal}
  \state{QC}
  \country{Canada}
}
\email{bmitra@microsoft.com}

\renewcommand{\shortauthors}{Yifan He, To Eun Kim, Fernando Diaz, Jaime Arguello, \& Bhaskar Mitra}

\newcommand{\movie}{Movie\xspace}
\newcommand{\landmark}{Landmark\xspace}
\newcommand{\person}{Person\xspace}

\newcommand{\gptfouro}{GPT-4o\xspace}
\newcommand{\gptfouromini}{GPT-4o-mini\xspace}

\begin{abstract}
Tip-of-the-tongue (TOT) search occurs when a user struggles to recall a specific identifier, such as a document title. While common, existing search systems often fail to effectively support TOT scenarios. Research on TOT retrieval is further constrained by the challenge of collecting queries, as current approaches rely heavily on community question-answering (CQA) websites, leading to labor-intensive evaluation and domain bias.
To overcome these limitations, we introduce two methods for eliciting TOT queries—leveraging large language models (LLMs) and human participants—to facilitate simulated evaluations of TOT retrieval systems. Our LLM-based TOT user simulator generates synthetic TOT queries at scale, achieving high correlations with how CQA-based TOT queries rank TOT retrieval systems when tested in the Movie domain. Additionally, these synthetic queries exhibit high linguistic similarity to CQA-derived queries. For human-elicited queries, we developed an interface that uses visual stimuli to place participants in a TOT state, enabling the collection of natural queries. In the Movie domain, system rank correlation and linguistic similarity analyses confirm that human-elicited queries are both effective and closely resemble CQA-based queries.
These approaches reduce reliance on CQA-based data collection while expanding coverage to underrepresented domains, such as Landmark and Person. LLM-elicited queries for the Movie, Landmark, and Person domains have been released as test queries in the TREC 2024 TOT track, with human-elicited queries scheduled for inclusion in the TREC 2025 TOT track. Additionally, we provide source code for synthetic query generation and the human query collection interface, along with curated visual stimuli used for eliciting TOT queries.
\end{abstract}
\maketitle
\section{Introduction}
\label{sec:introduction}

% 0-1. What is TOT
Tip-of-the-Tongue (TOT) retrieval is a known-item search task where the searcher aims to re-find a previously encountered item but struggles to formalize the query due to an inability to recall specific identifiers, such as a document title or a person’s name \cite{arguello-movie-identification, Bhargav-2022-wsdm}. 
Unlike other known-item retrieval tasks, 
% where the searcher provides a limited but concrete description of the target item, 
TOT information requests are typically verbose and complex, often blending semantic details about the context in which the searcher previously engaged with the item \cite{Meier21-complex-reddit, lee2006known}. 
These queries frequently include linguistic phenomena such as uncertainty, exclusion criteria, relative comparisons, and even false memories \cite{arguello-movie-identification, Meier21-complex-reddit}, making them especially challenging for traditional retrieval systems \cite{arguello-movie-identification, Bhargav-2022-wsdm, lin-etal-2023-whatsthatbook, arguello2023overview}.

% 0-3. WHY IS IT IMPORTANT
Addressing the challenges of TOT retrieval is critical because searchers in the TOT state experience higher level of frustration compared to other types of memory-related search failures \cite{elsweiler2007towards}.
TOT retrieval spans a wide range of domains, from casual leisure search \cite{elsweiler2011casualleisure}, such as finding games \cite{gameTOT}, music \cite{Bhargav23MusicTOT}, and books \cite{Bhargav-2022-wsdm}, to critical applications like email retrieval \cite{Elsweiler08emailrefinding, Elsweiler2011Seeding, Kim09desktop} and enterprise content search \cite{Dumais03enterprise}. In professional settings, ineffective TOT support can lead to significant economic costs, as employees spend excessive time retrieving essential information, reducing productivity and efficiency \cite{white2015enterprisesearch}.

% New version that covers point 1, 2, 3 %%%%%%%%%%%%
The evaluation of TOT retrieval systems faces significant challenges due to limitations in test collection resources and current dataset construction methods. Most existing TOT queries in current datasets are sourced from community question answering (CQA) websites, which fail to fully capture the diversity of real-world TOT information needs \cite{Meier21-complex-reddit}. This reliance on CQA platforms introduces domain skewness, as existing datasets and studies are heavily biased toward casual leisure topics like movies and books \cite{elsweiler2011casualleisure, bogers2025exploring}, while underrepresenting areas such as people and landmarks. Even attempts to expand domain coverage \cite{Meier21-complex-reddit} have had limited success, leaving many domains underrepresented.
Additionally, the labor-intensive nature of dataset construction---which relies on manual annotation, coding, and refinement \cite{Bhargav-2022-wsdm, Bhargav23MusicTOT, arguello-movie-identification, Meier21-complex-reddit}---slows progress and restricts the ability to conduct large-scale and multi-domain evaluations. 
% restricting the applicability of these datasets and hindering the development of retrieval systems that can effectively handle diverse TOT queries.

Advancing TOT retrieval research requires overcoming these challenges, as the lack of diverse and representative datasets leaves retrieval systems poorly optimized for real-world scenarios, reducing their ability to handle complex and ambiguous queries effectively.

However, current approaches fail to address these issues due to data scarcity, limited scalability, and restricted scope. Search engine query logs, which could provide real-world insights, remain largely inaccessible due to corporate privacy \cite{barbaro2006face} and confidentiality \cite{Poblete10confidentiality} concerns. 
As a result, prior efforts have relied on CQA platforms and manual labeling, but data collection restrictions from sites like Reddit, along with the scalability and domain bias issues, further limit dataset diversity and representativeness. These challenges underscore the need for more scalable and comprehensive approaches to TOT data collection and evaluation.

% 4. RESEARCH QUESTIONS
In response, this paper presents a novel evaluation framework for TOT retrieval systems, overcoming the limitations of CQA-based datasets by eliciting TOT queries from both large language models (LLMs) and humans. We develop an LLM-based TOT user simulator and a human query collection interface to address data scarcity, domain skewness, and scalability challenges, enabling more comprehensive evaluations.

Our research questions are as follows:\\
\textbf{RQ 1: Can we elicit TOT queries from LLMs for effective simulated evaluation of TOT retrieval systems?}
Unlike prior work, we explore LLM-elicited queries for TOT evaluation, validating our simulator by measuring system rank correlation and linguistic similarity with CQA-collected queries. By testing various prompt configurations and temperatures, we demonstrate that LLM-elicited queries can effectively support TOT retrieval evaluation with high validity.

\textbf{RQ 2: Can we elicit TOT queries from humans for effective simulated evaluation of TOT retrieval systems?}
To our knowledge, human-elicited TOT queries have not been previously used for evaluation. Similar to RQ1, we validate their effectiveness by comparing system rankings derived from human-elicited and CQA-based queries, demonstrating high alignment. Additionally, linguistic similarity analysis confirms that human-elicited queries share key characteristics with CQA-based queries, supporting their viability for TOT retrieval evaluation.

\textbf{RQ 3: Can the elicitation-based evaluation methods be used in other domains underrepresented in CQA-collected test collections?}
While the development and validation of our methods focus on the Movie domain---where ample CQA-collected data exists---we test their applicability to Landmark and Person domains, which are less represented in existing datasets. Validation results show that elicited queries can achieve high system rank correlations in these new domains, suggesting the potential research direction on the general method of collecting TOT queries across multiple domains.

In addition to the research contributions, the core outcomes of this work lie in its resource contributions:
\begin{itemize}
    \item A dataset of 450 synthetically generated TOT queries across three domains: Movie, Landmark, and Person.
    
    \item A method and prompt design for synthetically generating effective TOT queries from LLMs for simulated evaluation.\footnote{\url{https://github.com/kimdanny/llm-tot-query-elicitation}}
    
    \item A human TOT query elicitation interface designed to collect effective TOT queries for simulated evaluation along with a carefully curated visual stimuli collection, covering the Movie, Landmark, and Person domains.\footnote{\url{https://github.com/kimdanny/human-tot-query-elicitation-mturk}}
\end{itemize}

\section{Related Work}
%%%%%%%%%%%%%%
% Know-Item Retrieval and Query Simulation
%%%%%%%%%%%%%%
\subsection{Query Simulation \& Known-Item Retrieval}

Query simulation methods have been used for various purposes, including document expansion \cite{nogueira2019docT5query} and synthetic test collection generation \cite{Rahmani24synthetic}. In the context of known-item retrieval, these methods have been explored to improve retrieval strategies \cite{OgilvieCallan03combining, yalenlp-trec2024} and evaluation frameworks \cite{Azzopardi06testbeds, hagen2015corpus}.

%% Query Simulation
\textit{Simulating} the known-item queries has long been an active research area \cite{balog2006overviewWebclef, Azzopardi07SimulatedQueries, Kim09desktop, Elsweiler2011Seeding}.
Early work \cite{Azzopardi07SimulatedQueries} generated synthetic queries using term-based likelihood models, selecting query terms based on their likelihood within a randomly chosen document. Later studies adapted this approach for desktop search \cite{Kim09desktop} and email re-finding \cite{Elsweiler2011Seeding}, demonstrating its effectiveness for simulated evaluations of know-item retrieval models.
The \textit{validation} of these query simulators has also been a key focus.
System ranking correlation \cite{balog2006overviewWebclef}, retrieval score distribution comparisons \cite{Azzopardi07SimulatedQueries}, and synthetic versus human query resemblance \cite{Kim09desktop} have been used to assess their reliability.

While valuable, known-item search queries differ significantly from TOT queries, which are longer and more complex. Despite progress in simulating known-item queries, TOT retrieval remains unexplored. This paper bridges that gap by introducing novel TOT query elicitation methods and adapting established validation techniques \cite{zeigler2000theory} to ensure alignment with real-world queries, enabling scalable and accurate simulated evaluations.

%%%%%%%%%%%%%%
% TOT Datasets
%%%%%%%%%%%%%%
\subsection{TOT Datasets}
Several datasets have been developed to support research on TOT retrieval, primarily collected from online CQA platforms and focused on specific domains. MS-TOT \cite{arguello-movie-identification} was constructed from the \textit{IRememberThisMovie} website and human-annotated with tags in the Movie domain. It also includes qualitative coding of TOT queries and demonstrates significant room for improvement in current retrieval technologies for such information needs. Similarly, \citet{gameTOT} collected TOT queries from Reddit's \textit{/r/tipofmyjoystick} subreddit in the Game domain, providing coded tag information. Other datasets include Reddit-TOMT \cite{Bhargav-2022-wsdm}, focused on movies and books from Reddit's \textit{/r/tipofmytongue} subreddit; TOT-Music \cite{Bhargav23MusicTOT}, targeting the Music domain from the same subreddit; and Whatsthatbook \cite{lin-etal-2023-whatsthatbook}, sourced from \textit{GoodReads}, focused on the Book domain.

In response to the domain specificity of these datasets, recent efforts have aimed to expand TOT datasets across multiple areas. \citet{Meier21-complex-reddit} expanded to general casual leisure domains using data from six Reddit subreddits, including games, books, and music, although other identified domains, such as videos and people, remain underrepresented. Similarly, TOMT-KIS \cite{frobe2023-performance-pred} extended the collection from \textit{/r/tipofmytongue} by adapting \citet{Bhargav-2022-wsdm}'s approach with fewer filtering restrictions, resulting in 1.28 million TOT queries. However, only 47\% of these queries have identified answers, and the dataset continues to exhibit severe domain skewness toward a few topics.

In this work, we develop and validate TOT query elicitation methods using the Movie domain for robust evaluation, then expand to Landmark and Person to assess applicability across underrepresented domains.

\section{Method}
Overall, we elicit TOT queries from both LLMs and humans and validate them using two methods: system rank correlation (\S\ref{subsubsec:sys-rank-correlation}) and linguistic similarity (\S\ref{subsubsec:ling-sim}).

\subsection{Query Elicitation}
For LLM-elicited queries, we generate synthetic queries by exploring various prompting strategies. We experiment with different prompting conditions and model hyperparameters to identify the most effective prompt that yields the best validation results.  
This procedure is detailed in Section \S\ref{sec:llm-elicitation}.  

For human-elicited queries, we designed an interface that places participants in a TOT state using visual stimuli, such as movie stills, landmarks, and celebrity images. Participants then compose TOT queries as they would when posting on a CQA website.  
This procedure is described in Section \S\ref{sec:human-elicitation}.

\subsection{Query Validation}
% With the elicited queries, we perform two types of validation.

\subsubsection{\textbf{System Rank Correlation}}\label{subsubsec:sys-rank-correlation}
\begin{figure} 
\centering
\includegraphics[trim=140 130 90 120, clip, width=\columnwidth]{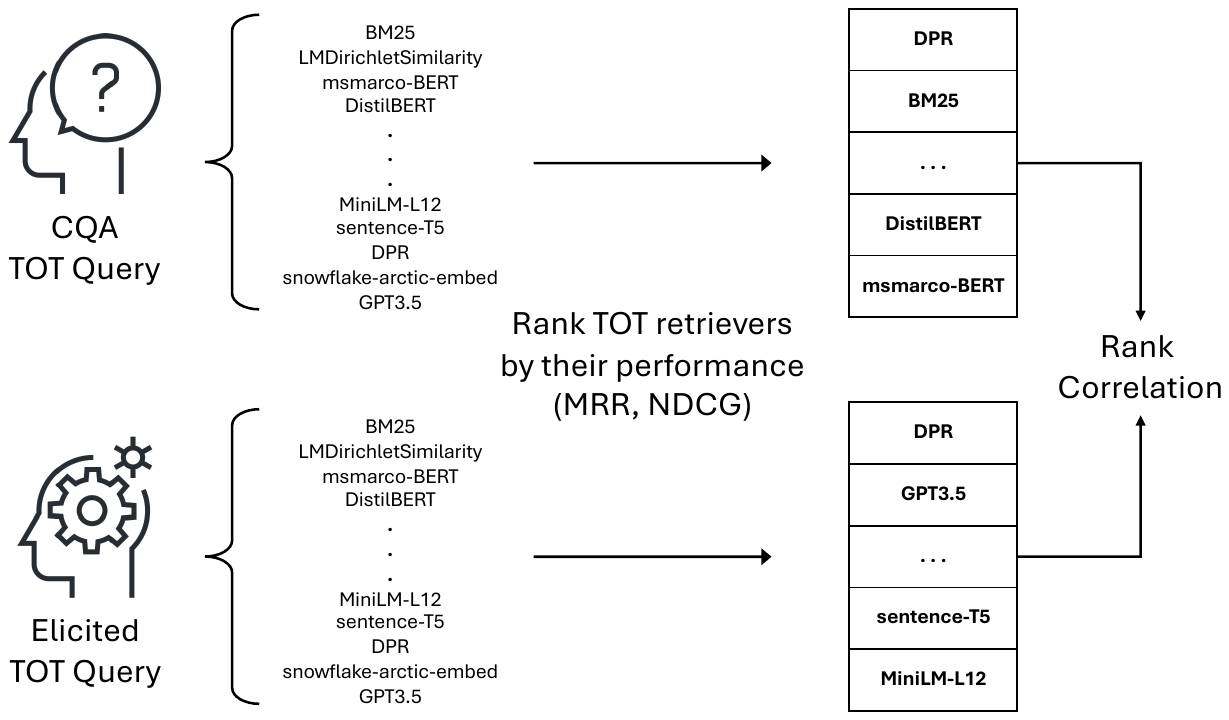}
\caption{
Validation of elicited queries using system rank correlation. We evaluate 40 different retrieval models using both CQA-based and elicited queries, ranking them based on search performance measured by MRR@1000 and NDCG@1000. We then compute Kendall’s Tau and Pearson correlation to assess the agreement between the rankings derived from the two query sets.
}
\label{fig:validation-sys-rank}
\end{figure}

To validate the effectiveness of elicited TOT queries, we measure the correlation between its rankings of retrieval systems and rankings based on CQA queries for the same entities. This evaluation assesses whether retrieval models maintain consistent performance across different query sources but on the same entities. If the rankings derived from elicited queries strongly correlate with those from CQA-based queries, it indicates that the elicited queries capture similar retrieval challenges and retrieval effectiveness. A high correlation suggests that our synthetic and human-elicited queries can serve as reliable substitutes for traditionally collected CQA-based queries in evaluating retrieval systems.

To compute these correlations, we run the queries on 40 different retrieval models, comprising lexical and dense retrievers. The lexical retrievers include BM25 \cite{Robertson1995OkapiBM25} and language models with Dirichlet priors \cite{zhai2001DirichletSmoothing} using varying parameters. The dense retrievers include models of different sizes and performance levels, such as MiniLM-L6 and MiniLM-L12 \cite{miniLM}. To further introduce variation in systems, we include retrieval models with intentionally degraded performance by reinitializing the weights of certain layers in dense retrievers. Additionally, we incorporate an API-based closed-source LLM as one of the retrieval systems, specifically GPT-3.5-Turbo-Instruct, following the prompting format used in the baseline of the TREC 2024 TOT track to function as a ranker.
Furthermore, we include the top-ranked retrieval system from the TREC 2023 TOT track \cite{luis24totDPR, arguello2023overview} to provide a strong performance reference point.

Figure \ref{fig:validation-sys-rank} illustrates our validation strategy, showing how retrieval system rankings across different query sets provide insight into the reliability and effectiveness of our elicited queries.

%%%%%%%%%%%%%%%%%%%%%%%%
\subsubsection{\textbf{Linguistic Similarity}}\label{subsubsec:ling-sim}

\begin{figure} 
\centering
\includegraphics[width=0.9\columnwidth]{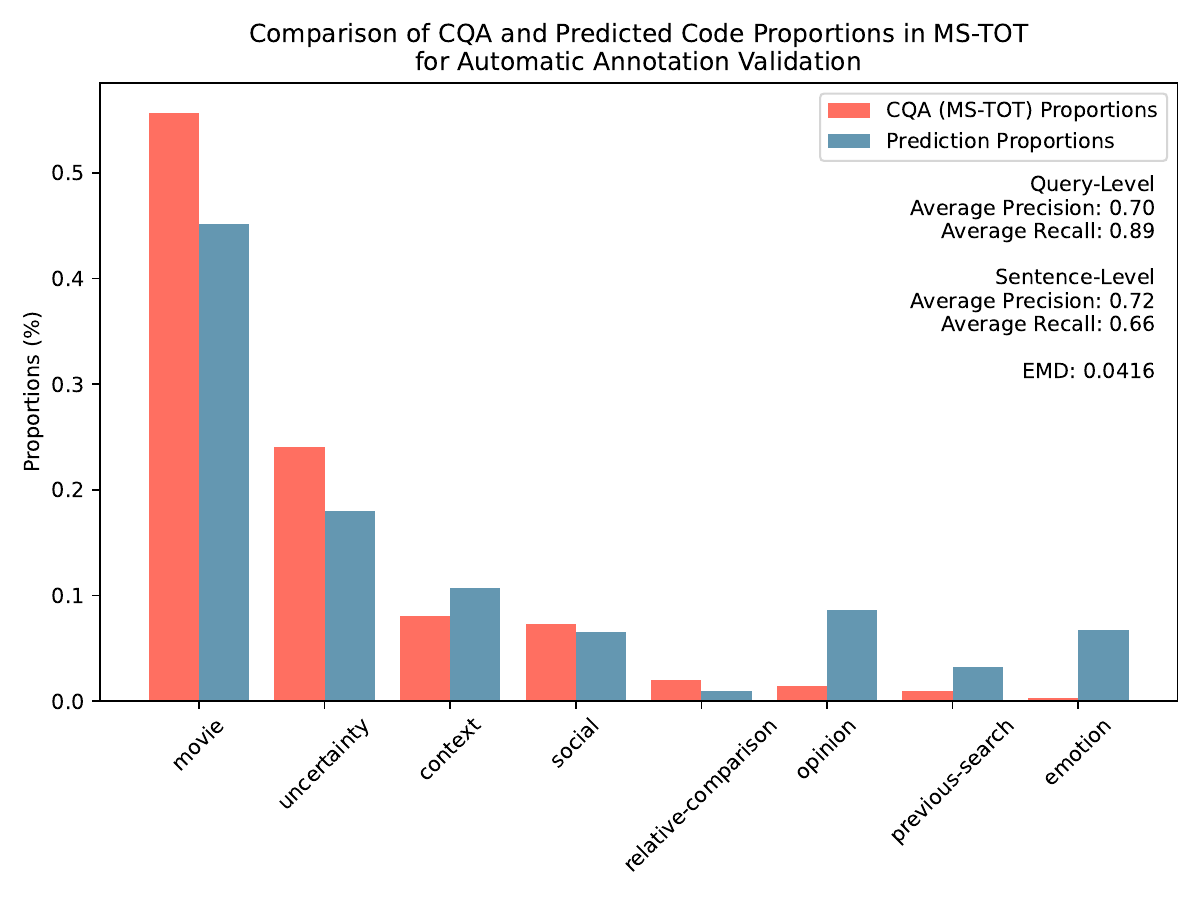}
\caption{
Validation of automatic annotation using the MS-TOT dataset.
Results show that GPT-4o-mini performs well in annotating TOT queries, closely aligning with human annotators. It achieves high accuracy and exhibits low Earth Mover’s Distance (EMD), indicating strong agreement with expert annotations.}
\label{fig:automatic-annotation-validation}
\end{figure}

Elicited TOT queries can exhibit substantial variation in writing style, word choice, experiences, and the presence of distorted memories \cite{Meier21-complex-reddit, arguello-movie-identification}. This diversity is inherent to the nature of TOT queries and is a valid characteristic of real-world TOT retrieval scenarios. Consequently, evaluating linguistic similarity between CQA-based and elicited queries using traditional methods, such as vector-based semantic similarity or lexicon-based similarity, may be ineffective.

To address this, we utilize a predefined set of TOT-specific linguistic codes introduced by \citet{arguello-movie-identification}. These handcrafted codes provide sentence-level annotations of TOT queries in the Movie domain, categorizing linguistic phenomena into eight top-level groups: `movie', `context', `previous-search', `social', `uncertainty', `opinion', `emotion', and `relative-comparison'. By leveraging this framework, we compare the linguistic distribution of codes between CQA-based and elicited queries rather than relying on direct semantic or lexical overlap.

To conduct this analysis, we annotate our elicited TOT queries in the Movie domain using these linguistic codes and compare the percentage distributions of codes found in CQA-based and elicited queries. To automate this process, we develop a language model-based automatic code annotator, prompting GPT-4o-mini to produce JSON-formatted sentence-level annotations.\footnote{GPT model's temperature is set to 0 for reproducibility and consistency.}

Before applying this annotator to LLM- and human-elicited queries, we first validate its performance on the MS-TOT dataset, where sentence-level gold annotations are available. We evaluate the annotator’s performance by computing precision and recall as prediction accuracy measures. Additionally, to assess annotation quality at a broader level, we compute query-level precision and recall, which measure the accuracy of identifying unique codes that appear within a multi-sentence query.

To further validate the annotator, we compare the distribution of annotated codes against the gold annotations using Earth Mover’s Distance (EMD), which quantifies how different the predicted code distribution is from the reference distribution. Figure \ref{fig:automatic-annotation-validation} presents the validation results of our automatic annotator on the MS-TOT dataset. Our evaluation shows that the annotator achieves sufficiently high accuracy and low EMD, confirming its suitability as an automatic labeler.

With this validation, we apply the automatic annotator to both LLM- and human-elicited queries in the Movie domain and analyze how their linguistic code distributions compare to those in CQA-based queries. However, we conduct this linguistic similarity validation only in the Movie domain, as there are no existing comprehensive handcrafted linguistic codes available for the Landmark and Person domains.

\section{TOT Query Elicitation from LLMs}\label{sec:llm-elicitation}

\subsection{Synthetic Query Generation}
\begin{table*}[]
\centering
\begin{tabular}{ccccccc c ccc}
   &&&&&&&\multicolumn{3}{c}{Validation Results}\\
   \cline{8-10}
   & \multicolumn{4}{c}{Prompt \& Model Design}   &&& \multicolumn{1}{c}{MRR-based} & \multicolumn{1}{c}{NDCG-based} & \multicolumn{1}{c}{\makecell{Linguistic\\Similarity}} \\
   \cline{2-6} \cline{8-10}
   ID & \makecell{Prompt\\Template} & \makecell{System\\Role}   & \makecell{Wikipedia\\Summary} & \makecell{Instruction\\Type} & Temperature && $\tau$/$r$ $\uparrow$ & $\tau$/$r$ $\uparrow$ & EMD $\downarrow$ \\
   \hline
1 & V0 & Writing Assistant  & w/o  & 9 rules                 & 0.5  && 0.2641 / 0.6101 & 0.3092 / 0.4247 & 0.0899  \\
2 & V1 & Writing Assistant  & w/   & 9 rules                 & 0.5  && 0.4923 / 0.5692 & 0.5077 / 0.6475 & 0.1012  \\
3 & V2 & Searcher role play  & w/   & 6-shot                  & 0.5  && 0.4632 / 0.5391 & 0.4430 / 0.5492 & 0.0299  \\
4 & V2 & Searcher role play  & w/   & 6-shot                  & 0.7  && 0.4538 / 0.4234 & 0.4310 / 0.4650 & 0.0276  \\
5 & V3 & Searcher role play  & w/o  & 6-shot                  & 0.5  && 0.1524 / 0.5293 & 0.2045 / 0.5995 & 0.0584  \\
6 & V3 & Searcher role play  & w/o  & 6-shot                  & 0.7  && 0.1440*/ 0.5173 & 0.1960*/ 0.5942 & 0.0534  \\
7 & V4 & Searcher role play & w/   & 13 rules                & 0.3  && 0.6927 / 0.9143 & 0.6757 / 0.8975 & 0.0490  \\
8 & V4 & Searcher role play & w/   & 13 rules                & 0.5  && 0.6463 / 0.9173 & 0.6235 / 0.8998 & 0.0542  \\
9 & V4 & Searcher role play & w/   & 13 rules                & 0.7  && 0.6472 / 0.9336 & 0.6785 / 0.9303 & 0.0577  \\
10 & V5 & Searcher role play & w/   & 14 rules               & 0.1 && 0.6719 / 0.9388 & 0.6500 / 0.9375 & 0.0542   \\
11 & V5 & Searcher role play & w/   & 14 rules               & 0.3 && 0.6558 / 0.9396 & 0.6728 / 0.9342 & 0.0620   \\
12 & V5 & Searcher role play & w/   & 14 rules               & 0.5 && 0.7013 / \textbf{0.9536} & 0.7127 / \textbf{0.9500} & 0.0590   \\
13 & V6 & Searcher role play & w/   & 7 Musts + 7 Coulds     & 0.3 && \textbf{0.7573} / 0.9166 & \textbf{0.7194} / 0.8973 & \textbf{0.0264} 
\end{tabular}
\caption{
Experiments on LLM-elicited query generation in the Movie domain.
$\tau$ and $r$ represent Kendall's Tau and Pearson's r correlation values, respectively, while EMD (Earth Mover’s Distance) measures linguistic similarity between CQA-based and LLM-elicited queries. All correlation values have a p-value < 0.01, except for those marked with (*).
}
\label{tab:prompt-versions}
\end{table*}

\begin{figure} 
\centering
\includegraphics[trim=110 243 78 210, clip, width=\columnwidth]{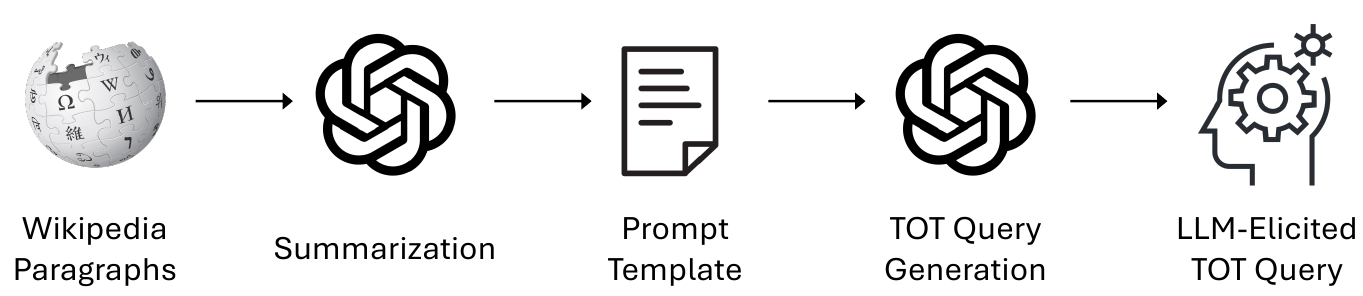}
\caption{
LLM-elicited TOT query generation process. Given an arbitrary entity for which we want to generate a TOT query, we retrieve its Wikipedia page, summarize it into a few paragraphs, and construct a prompt template using the summary. This prompt is then used to query an LLM to generate a synthetic TOT query.
}
\label{fig:query-gen-process}
\end{figure}

Figure \ref{fig:query-gen-process} illustrates the LLM-elicited TOT query generation process. Given an arbitrary entity for which we want to generate a TOT query, we retrieve the entity’s Wikipedia page, summarize it into a few paragraphs, and construct a prompt template using the summary. This prompt is then used to query an LLM to generate a synthetic TOT query.

The query generation process consists of the following steps:
\begin{enumerate}
\item \textit{Sampling a target entity}:\\
A random entity is selected as the target for TOT query generation. In the Movie domain, the MS-TOT test set serves as the pool of entities, while for the Landmark and Person domains, we manually curated the entity pool.

\item \textit{Retrieving and summarizing the entity’s Wikipedia page}:\\
The entity’s Wikipedia page is retrieved and used to construct a summarization prompt for the GPT-4o model.\footnote{GPT4o-2024-05-13} To ensure the content fits within the model’s token limit, the page is truncated if necessary. The model processes the page into a two-paragraph summary. For Movie domain entities, we check for a dedicated \textit{Plot} section and ensure its content is included in the summarization prompt, as it often provides essential context for generating realistic TOT queries.

\item \textit{Constructing a domain- and entity-specific prompt}:\\
Following summarization, a domain- and entity-specific prompt is constructed, incorporating the summarized information to generate a TOT query. The same GPT-4o model is used for query elicitation. 
We tested various prompt configurations and temperature settings to identify the best-performing strategy in the Movie domain, then applied it to the Landmark and Person domains.\footnote{Details on the different prompt versions we tested are available at \url{https://github.com/kimdanny/llm-tot-query-elicitation}.}

\item \textit{Ensuring entity name anonymity in the generated query}:\\
An essential consideration in the query generation process is ensuring that the generated TOT query does not contain the target entity’s name. To address this, we implement a name-checking step: if the generated query includes the entity name, we retry the generation using the same prompt, with a maximum of three retries before discarding the query.
\end{enumerate}

%%%%%%%%%%
% Results
%%%%%%%%%%
\subsection{Results and Analyses}
\textbf{RQ 1: Can we elicit TOT queries from LLMs for effective simulated evaluation of TOT retrieval systems?}

We tested various configurations to identify the most effective TOT query generation approach. The final prompt structure was determined through the extensive testing of different prompt templates, system roles, instruction types, and temperature settings, as well as evaluating the inclusion of Wikipedia summaries (Table \ref{tab:prompt-versions}). 
For each configuration, queries were generated using entities from the test set of the TREC TOT Track 2023 \cite{arguello2023overview}, derived from the MS-TOT dataset \cite{arguello-movie-identification}, and evaluated based on system rank correlation and linguistic similarity.

The top-performing strategies (Experiment ID 12 and 13) utilized a role-playing scenario, where the LLM assumed the role of a TOT searcher. The prompt began with:
\begin{tcolorbox}
Let's do a role play. You are now a person who watched a movie \{TOT Entity\} a long time ago and forgot the movie's name. 
\end{tcolorbox}
\noindent
This setup encouraged the LLM to generate queries that better captured real TOT search behavior, leading to improved retrieval rankings.

Specifically, the prompt included a summarized Wikipedia entry followed by 14 guideline rules, ensuring the searcher role-play aligned with real user behavior. 
For example, rules such as \textit{"Reflect the imperfect nature of memory with phrases that express doubt or mixed recollections"} and \textit{"Introduce a few incorrect or mixed-up details to make the recollection seem more realistic and challenging to pinpoint"} helped simulate the uncertainty and distorted memory typical of TOT searchers.
Similarly, \textit{"Share a personal anecdote related to when or with whom you watched the movie"} and \textit{"Focus on sensory details such as the overall mood, sounds, or emotional impact of the movie, using vivid descriptions"} aimed to incorporate the personal and emotional elements often found in real TOT queries.

While Experiment 12 achieved the highest Pearson correlation, Experiment 13 
resulted in the highest Kendall’s Tau correlation and the lowest linguistic distance from CQA-based queries.
A notable comparison is between Experiment IDs 11 and 13, which share the same temperature and 14 rules. The key difference lies in the rule structuring: while ID 11 presented all 14 rules equally, ID 13 divided them into 7 ``must" rules and 7 ``could" rules. This slight difference led to a significant increase in Kendall’s Tau and linguistic similarity, despite a minor drop in Pearson correlation.

Through the experiments of different setups, we derived several key insights. Role-playing as a TOT searcher with explicit behavioral guidelines led to more realistically structured TOT queries. Providing generation rules proved more effective than few-shot prompting, as seen when transitioning from Prompt Version 2 (few-shot) to Prompt Version 4 (rule-based), which significantly improved validation scores.

Including a Wikipedia summary in the prompt substantially increased system rank correlation, indicating that structured background knowledge helps the LLM generate queries more aligned with real TOT queries. Additionally, lower temperature settings (0.3, 0.5) outperformed higher ones (0.7), contrary to our initial hypothesis that increased randomness would better simulate the stochastic nature of memory retrieval. The improved performance at lower temperatures likely stems from the model following the defined generation rules more closely.

\begin{figure} 
\centering
\includegraphics[width=0.9\columnwidth]{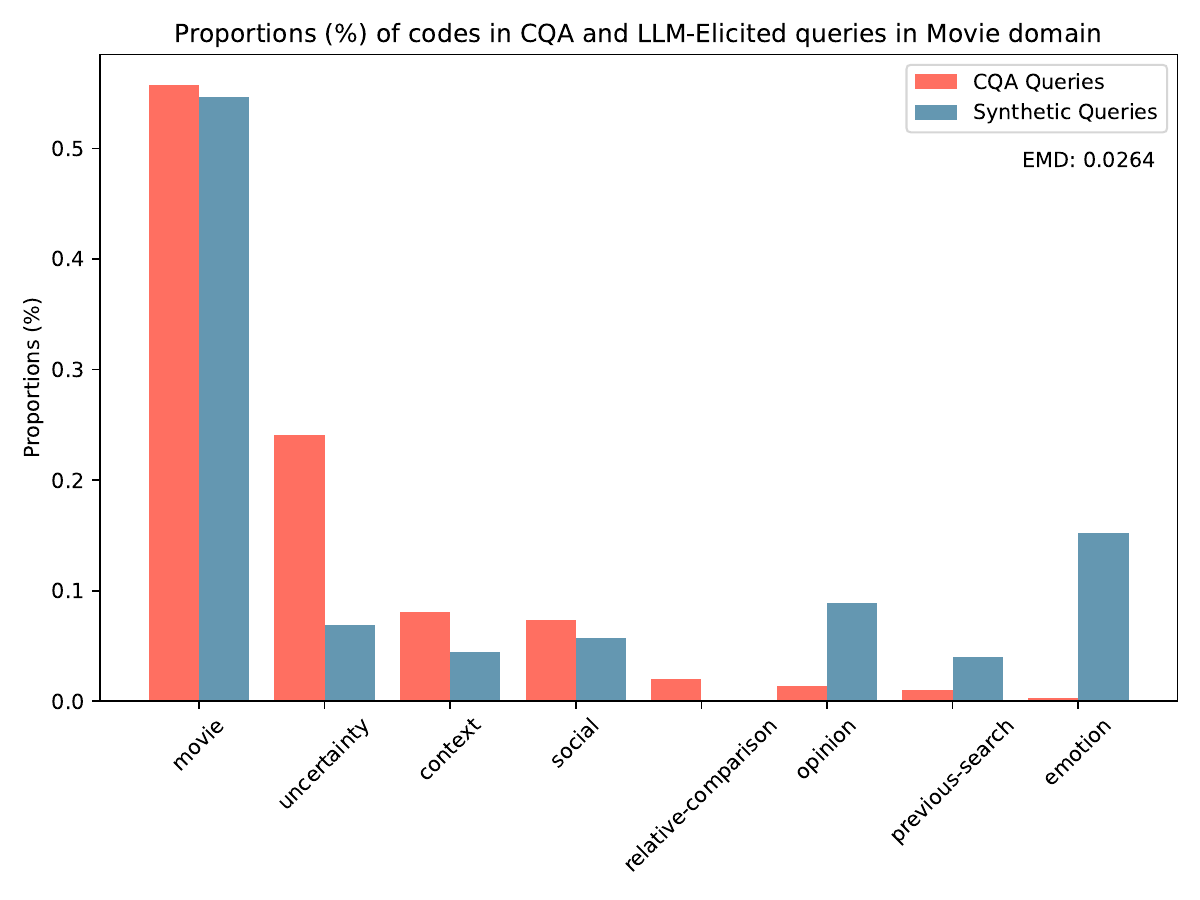}
\caption{
Proportions of labeled codes in CQA and LLM-elicited queries in the Movie domain using Prompt Version V6 (Experiment ID 13). This setup achieved the lowest frequency distribution difference (EMD) among all experiments.
}
\label{fig:cqa-llm-v6-ling}
\end{figure}

Additionally, linguistic similarity analysis using Earth Mover’s Distance (EMD) resulted in a value of 0.0264 (Figure \ref{fig:cqa-llm-v6-ling}) in Experiment 13, showing the lowest value. 

Based on findings across the two validation methods, we empirically conclude that LLM-generated TOT queries are an effective approach for designing simulated evaluations of TOT retrieval systems.

\textbf{RQ 3-1: Can the LLM-elicitation-based evaluation methods be used to other domains underrepresented in CQA-collected test collections?}

Based on our findings, we adopted the Experiment ID 13 configuration for the final dataset generation, extending its application to the underrepresented Landmark and Person domains. To adapt the Prompt V6 setup for these domains, we adjusted entity descriptions while preserving the core prompt structure.
For instance, the rule \textit{"Share a personal anecdote related to when or with whom you watched the movie"} was modified to \textit{"Share a personal anecdote related to the person"} to align with the context of the Person domain.

To validate the generated queries, we compared them against CQA-collected queries. Since existing TOT test collections are largely focused on the Movie domain, no established test collection exists for the Landmark and Person domains. To address this, we hand-curated a test collection, resulting in 22 Landmark-domain queries and 70 Person-domain queries from the Reddit's subreddit, called \textit{/r/tipofmytongue}.

\begin{table}[]
\centering
\begin{tabular}{c|cc}
   % &&&&&&&\multicolumn{3}{c}{Validation Results}\\
   % \cline{8-10}
   & \multicolumn{1}{c}{MRR-based} & \multicolumn{1}{c}{NDCG-based} \\
   % \cline{2} \cline{3}
   Domain & $\tau$/$r$ $\uparrow$ & $\tau$/$r$ $\uparrow$ \\
   \hline
   Landmark & 0.5984 / 0.8227 & 0.6967 / 0.9142  \\
Person & 0.6362 / 0.7231 & 0.5691 / 0.7996  
\end{tabular}
\caption{
Kendall's Tau and Pearson's r correlation values between CQA-based queries and LLM-elicited queries in the Person and Landmark domains.
$\tau$ and $r$ represent Kendall's Tau and Pearson's r correlation values, respectively. All correlation values have a p-value < 0.01.
}
\label{tab:person-landmark-correlation}
\vspace{-10pt}
\end{table}

Table \ref{tab:person-landmark-correlation} presents the system rank correlation validation results for both domains. While the validation scores are not as strong as those observed in the Movie domain, they still demonstrate reasonably high values. This suggests that LLM-elicited query generation methods can be extended to underrepresented domains and could potentially be developed into generalizable LLM-elicitation techniques across multiple domains.

\section{TOT Query Elicitation from Human}\label{sec:human-elicitation}

The study of the tip-of-the-tongue phenomenon has been a long-standing area of research in psychology \cite{burke1991tip, jones1989back}. Many studies have attempted to induce TOT states in participants using auditory \cite{reefer1995name} or visual stimuli \cite{tranel2005landmarks}, enabling researchers to examine recognizability (whether a subject recognizes an entity from the stimulus) and retrievability (whether they can recall the entity's name or title).

Building on these methodologies, we employ visual stimuli in the Movie, Landmark, and Person domains to develop an interface that places participants in a TOT state and allows them to compose TOT queries about the entities they struggle to recall.

In this section, we describe the design process of our interface for eliciting human-written TOT queries from trained contracted participants, along with an analysis of the collected human-elicited queries.

\subsection{Visual Stimuli Collection}

To obtain images in Movie, Landmark, and Person domains, we collected entities from Wikipedia, selecting pages based on their infobox namespace and the presence of images.\footnote{List of infoboxes: \url{https://en.wikipedia.org/wiki/Wikipedia:List_of_infoboxes}.}
For the Person and Landmark domains, we directly collected images from Wikipedia pages. However, for the Movie domain, we identified linked IMDb or TMDB IDs on the Wikipedia pages and retrieved high-quality backdrop or still-cut images from TMDB. This approach was necessary because Wikipedia’s movie images were predominantly posters, which often contain the movie title and cannot be used to elicit a TOT state. TMDB backdrops, on the other hand, provide spoiler-free, high-quality visuals suitable for this purpose.\footnote{\url{https://www.themoviedb.org/bible/image/}}
For the Landmark domain, we selected ``Place" as a top-level category, and for the Person domain, we chose ``Person".

Selecting appropriate infobox namespaces within these top-level categories presented challenges, as the collected images needed to be \textit{unambiguous} and \textit{popular} enough for human participants to recognize the entities. For example, in Landmark domain, images of random parks often depicted generic green spaces, making them ambiguous, whereas iconic parks were more likely to be recognizable. 
For the Person domain, ambiguity was defined as the presence of multiple people in the image, which could confuse participants.
To ensure the quality of the stimuli, two authors rated infobox namespaces under the ``Place" and ``Person" categories on a three-point scale:
\begin{itemize}
    \item 0: Most images would be ambiguous or unrecognizable.
    \item 1: Some images might be ambiguous.
    \item 2: Most images would be unambiguous and recognizable.
\end{itemize}
Given that our participants were based in the United States, this factor was also considered during the labeling process. 

After completing the rating and ranking process, we selected one infobox namespace (`film') for Movie, 94 for Landmark (e.g., `amusement park', `bridge', `ancient site'), and 123 for Person (e.g., `comedian', `sportsperson', `academic').

Next, we collected Wikipedia pages associated with the selected infobox namespaces and sorted them by page view counts, using the Wikipedia database dump from October 1, 2023.\footnote{\url{https://dumps.wikimedia.org/enwiki/}} We selected the top 20\% of pages by view count, as less popular entities were deemed unlikely to be recognized by participants.

Finally, we manually filtered the collected images to remove unsuitable content. For the Movie domain, we excluded inappropriate images, such as those featuring nudity or excessive gore, and posters containing the movie title. For the Landmark domain, we removed images where the entity name was visible, such as station names in train station pictures. For the Person domain, we excluded images with multiple individuals.\footnote{Detailed image deselection criteria can be found at \url{https://github.com/kimdanny/human-tot-query-elicitation-mturk/blob/main/image_deselection_criteria.md}.} After filtering, we finalized a visual stimuli collection of 1,687 Movie entities, 1,946 Person entities, and 330 Landmark entities, all associated with Wikipedia pages, ensuring a diverse and high-quality set of stimuli for eliciting TOT queries.

\begin{figure*}[]
\centering
\includegraphics[trim=0 160 0 120, clip, width=\textwidth]{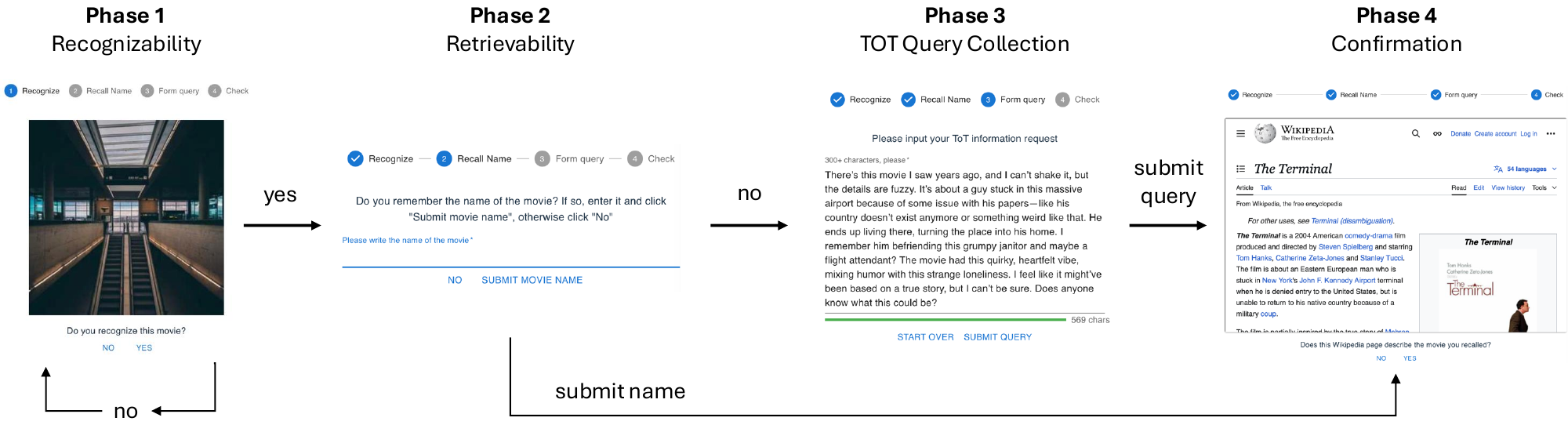}
\caption{
Flowchart of the Human TOT Query Elicitation Interface.
}
\label{fig:interface-flowchart}
\end{figure*}
\subsection{Interface Design}

We designed the flow of an interface as illustrated in Figure \ref{fig:interface-flowchart}.

\subsubsection{\textbf{Phase 1 (Recognizability)}}
In this phase, the system presents an image drawn from the pool of visual stimuli, asking participants whether they recognize the movie, landmark, or person depicted in the image, depending on the domain. The process repeats, drawing new images until the participant confirms recognition of an entity.

To ensure a balanced selection process, images in each domain were grouped into 20 buckets based on their popularity, measured by Wikipedia page view counts. When selecting an image, the system draws one random image per bucket in sequential order. The three domains---Movie, Landmark, and Person---are covered in a round-robin fashion, ensuring that images from different domains are presented in a balanced manner.

This approach minimizes the likelihood of a participant encountering the same image multiple times within a session and prevents from disproportionately drawing easily recognizable entities, leading to a more effective and controlled TOT induction process.

\subsubsection{\textbf{Phase 2 (Retrievability)}}
In this phase, we assess whether the participant can recall the name or title of the entity they recognized in Phase 1. If they cannot recall the name, it indicates that they are in a TOT state, and the interface transitions to Phase 3.

If they can recall the name, they are prompted to submit it, and the interface moves directly to Phase 4.

\subsubsection{\textbf{Phase 3 (TOT Query Collection)}}
If a participant reaches Phase 3, they are prompted to write a TOT query describing the entity they recognized but could not retrieve.

To ensure high-quality queries, we analyzed word count distributions in the MS-TOT dataset and determined that a minimum of 300 characters is necessary to effectively capture the complexity of TOT queries. To encourage longer responses, we followed the approach of \citet{agapie-13-longerqueries} and implemented a color-coded progress bar that visually indicates the current query length. The bar starts as red when the query is below 200 characters, turns yellow at 200 characters, and finally turns green once it reaches 300+ characters, signaling that the recommended length has been met.

Once the participant submits their TOT query, the interface transitions to Phase 4.

\subsubsection{\textbf{Phase 4 (Entity Confirmation)}}

In Phase 4, the system displays the Wikipedia page of the entity, allowing the participant to verify whether the entity they wrote a TOT query about or successfully retrieved is correct. This confirmation step is crucial for determining whether the collected TOT query is valid. 

After completing Phase 4, the interface loops back to Phase 1, where a new image is presented to continue the process.

\subsection{Results and Analyses}
\begin{table}[]
\centering
\resizebox{\columnwidth}{!}{%
\begin{tabular}{lrrrr}

\textbf{Phase}               & \textbf{Movie} & \textbf{Landmark} & \textbf{Person} & \textbf{Total} \\ \hline
Phase 1: Recognizability                &             &                & \\
\hspace{1em}Yes                & 672            & 231               & 1279               & 2182           \\ 
\hspace{1em}No                 & 5513           & 2079              & 5159               & 12751          \\ \hline
Phase 2: Retrievability                &             &                & \\
\hspace{1em}Yes                & 395            & 82                & 816                & 1293           \\ 
\hspace{1em}Yes (Correct)       & 314            & 23                & 721                & 1058           \\ 
\hspace{1em}Yes (Incorrect)     & 81             & 59                & 95                 & 235            \\ 
\hspace{1em}No                 & 273            & 149               & 462                & 884            \\ \hline
Phase 3: Query Collection                     & 245            & 119               & 413                & 777            \\ \hline
Phase 4: Entity Confirmation                &             &                & \\
\hspace{1em}Yes                & 178            & 57                & 349                & 584            \\ 
\hspace{1em}No                 & 61             & 61                & 59                 & 181            \\ 
\hspace{1em}N/A                & 6              & 1                 & 5                  & 12             \\ 
\end{tabular}
}
\caption{Statistics of human-elicited TOT query collection across different phases of the interface.}
\vspace{-10pt}
\label{tab:human-collection-stat}
\end{table}

Seven expert contracted participants interacted with the designed interface to generate human-elicited TOT queries.

Table \ref{tab:human-collection-stat} shows the number of instances recorded at each phase for the Movie, Landmark, and Person domains, along with their totals.
"Phase 1 - Yes" and "Phase 1 - No" indicate whether participants recognized the presented entity. "Phase 2 - Yes" represents cases where participants recalled the entity's name, with "Correct" and "Incorrect" indicating the accuracy of their recall. "Phase 2 - No" corresponds to participants who could not retrieve the entity's name, leading to TOT query collection in Phase 3. "Phase 4 - Yes" and "Phase 4 - No" denote whether the retrieved entity was correctly confirmed, while "Phase 4 - N/A" represents cases where confirmation was not applicable.\footnote{Due to session restarts or participants logging off, the numbers in "Phase 2 - No" and "Phase 3" do not always match exactly. Similarly, the sum of "Phase 2 - Yes" and "Phase 2 - No" does not always correspond to the number of "Phase 1 - Yes".}

From this human TOT query collection process, we obtained 178 Movie queries, 57 Landmark queries, and 349 Person queries, resulting in a total of 584 human-written TOT queries across three domains. 
The following is an example of a Landmark query:\\

\noindent\fbox{
\parbox{0.95\columnwidth}{
\textbf{Human-Elicited Landmark Query}:\\
i've seen a picture of this landmark but I can't quite remember where it is. it might have a garden area inside of it. you can walk from the bottom around and around to the top if i remember correctly. i don't think it's for housing. it's a sculpture. possibly located in new york or at least in the US.\\
\textbf{Correct Answer: Vessel}
}    
}\\

\noindent
As shown in the example, we qualitatively identified common linguistic patterns, such as expressions of uncertainty and distorted memories, within the collected queries.

\begin{table}[h]
\begin{tabular}{l|lll}
                & Movie & Landmark & Person \\ \hline
Recognizability & 0.11  & 0.10     & 0.20      \\
Retrievability  & 0.05  & 0.01     & 0.17      \\ 
\end{tabular}
\caption{Recognizability (Phase 1 - Yes / Phase 1 - Total) and Retrievability (Phase 2 - Yes Correct / Phase 1 - Total) in the human TOT query collection.}
\label{tab:psychology-connection}
\end{table}

Table \ref{tab:psychology-connection} presents the recognizability and retrievability of visual stimuli in our human experiments. Recognizability is measured as the ratio of participants recognizing an entity (Phase 1 - Yes) to the total instances in Phase 1. Retrievability is measured as the proportion of correctly recalled entity names (Phase 2 - Yes Correct) relative to all instances in Phase 1. These values provide insights into how well different domains induce TOT states in participants with our visual stimuli set.

\begin{figure} 
\centering
\includegraphics[width=0.9\columnwidth]{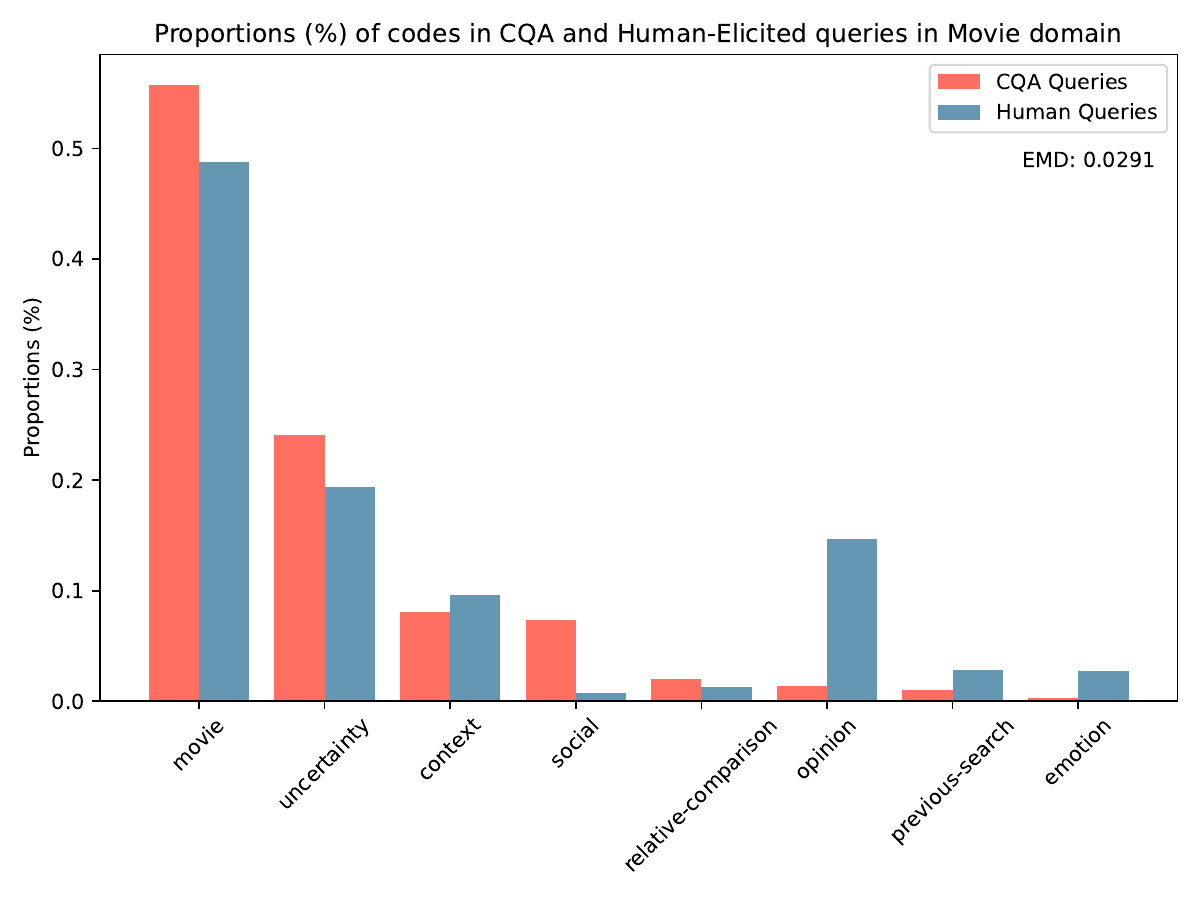}
\caption{
Proportions of labeled codes in CQA and human-elicited queries in the Movie domain, showing a low difference in frequency distribution (EMD).
}
\label{fig:cqa-human-ling}
\vspace{-10pt}
\end{figure}

To validate with system rank correlation, we needed overlapping entities between CQA data and our human-elicited TOT dataset. Since MS-TOT and manually collected Reddit posts had negligible overlap in the Landmark and Person domains, we turned to TOMT-KIS for matching entities.

As TOMT-KIS lacks explicit entity names, we matched entities by checking if the target name appeared as a substring in the answer posts. Only one valid match ("Fort Ticonderoga") was found in the Landmark domain, and nearly all matched queries in the Person domain were about movies or songs, as TOMT-KIS did not collect person-specific queries. Thus, we had to exclud both domains from validation.
For the Movie domain, we filtered TOMT-KIS queries for relevance by requiring the word "movie" while excluding "song", removing answers with multiple guesses, eliminating queries based solely on video links, and selecting the longest text for duplicate identifiers. This process yielded 303 Movie-domain queries, which we used for system rank correlation validation.

\textbf{RQ 2: Can we elicit TOT queries from humans for effective simulated evaluation of TOT retrieval systems?}
\begin{table}[]
\centering
\begin{tabular}{c|cc}
   & \multicolumn{1}{c}{MRR-based} & \multicolumn{1}{c}{NDCG-based} \\
   Domain & $\tau$/$r$ $\uparrow$ & $\tau$/$r$ $\uparrow$ \\
   \hline
   Movie & 0.6111 / 0.5882 & 0.6438 / 0.6889  \\
\end{tabular}
\caption{
Kendall's Tau and Pearson's r correlation values between CQA-collected queries (TOMT-KIS) and human-elicited queries in the Movie domain.
$\tau$ and $r$ represent Kendall's Tau and Pearson's r correlation values, respectively. All correlation values have a p-value < 0.01.
}
\label{tab:human-movie-correlation}
\vspace{-10pt}
\end{table}

Our results indicate that human-elicited TOT queries can serve as an effective resource for simulated evaluation of TOT retrieval systems. The system rank correlation between human-elicited queries and CQA-collected queries (TOMT-KIS) in the Movie domain shows high agreement (Table \ref{tab:human-movie-correlation}).
Additionally, linguistic similarity analysis using Earth Mover’s Distance (EMD) resulted in a value of 0.0291 (Figure \ref{fig:cqa-human-ling}), which is comparable to the lowest EMD achieved during the LLM-elicitation experiments. 

The higher proportion of opinion-related expressions in human-elicited queries may be influenced by differences in query intent. In CQA settings, users often structure their queries to maximize answerability, prioritizing factual details over personal reflections. In contrast, our elicitation method placed participants in a TOT state without the expectation of receiving an answer from community, which may have encouraged them to articulate their thoughts more freely. This introspective process could have naturally led to a greater inclusion of subjective impressions and emotions in their queries.

\textbf{RQ 3-2: Can the human-elicitation-based evaluation methods be used to other domains underrepresented in CQA-collected test collections?}

Since validation was restricted to the Movie domain due to the lack of sufficient CQA-collected queries in other domains, additional validation methods may be necessary to fully address the research question. However, the low EMD score (0.0291) in the Movie domain suggests that human-elicitation methods effectively capture authentic TOT queries, at least within this domain. Given that the elicitation process was designed to be domain-agnostic, follows established psychological principles of TOT states, and does not rely on movie-specific heuristics, these methods are likely to generalize to underrepresented domains, making them a promising direction for future TOT retrieval evaluations.

\section{Discussion}\label{sec:resource}

\textbf{Resource Contribution}.
Using the LLM-elicitation method from Experiment ID 13 (Prompt Version 6), we generated synthetic TOT queries for entities collected through the visual stimuli selection process, resulting in 1,687 queries in the Movie domain, 330 in the Landmark domain, and 1,946 in the Person domain. Additionally, through the human-elicitation method, we collected 584 human-elicited TOT queries spanning all three domains.

The full release of these queries is scheduled for the TREC 2025 TOT track, where they will be included as part of the official test collection. However, for TREC 2024 \cite{arguello2024overview}, and in this work, we have released 450 synthetic queries (150 per domain) from the full set of generated queries. Each query in the dataset is accompanied by its corresponding Wikidata ID, domain name, and entity name, ensuring clear entity association for retrieval experiments and analysis. Alongside these queries, we provide the source code for query generation and experimentation, as well as the visual stimuli entity set with corresponding image URLs and Wikidata ID. We also release the MTurk-based human query collection interface, allowing researchers to replicate or extend the human TOT query elicitation process.

\textbf{Availability}.
At the time of review, LLM-elicited queries are publicly available as part of the TREC 2024 TOT track \cite{arguello2024overview} test collection and can also be accessed at the track website.\footnote{\url{https://github.com/kimdanny/llm-tot-query-elicitation}} The human-elicited queries will be released as part of the TREC 2025 TOT track, aligning with the SIGIR 2025 conference.
Although the human-elicited queries are not yet publicly available, we have released the source code for the human query collection interface used in pilot testing on Amazon MTurk.\footnote{\url{https://github.com/kimdanny/human-tot-query-elicitation-mturk}} This allows researchers to explore and reproduce the query elicitation process for future development.

Both datasets are, and will continue to be, freely available under open licensing terms, ensuring unrestricted access for academic researchers and industry practitioners to support research and development in TOT retrieval.

\textbf{Utility}.
The resource is well-documented and designed for easy integration into retrieval experiments. Queries are provided in pure text format for compatibility with retrieval models, and a baseline implementation with tools for data loading and retrieval is available in the public repositories.
Additionally, this paper details the data provenance, processing, and experimentation steps, ensuring that future researchers can expand the dataset or adopt the TOT query elicitation method for other domains, supporting reproducibility and innovation.

\textbf{Novelty and Predicted Impact}.
Our work represents a major shift in TOT query collection methodology, moving beyond CQA-based datasets to LLM- and human-elicited queries, providing a scalable and flexible alternative for TOT dataset creation. Unlike previous approaches, our method eliminates the need for manual labeling, avoids data restrictions, and mitigates domain skewness in CQA datasets, which overrepresent casual leisure topics like movies and books. By incorporating underrepresented domains such as Person and Landmark, our dataset extends beyond traditional leisure-focused queries, supporting the development of general-domain TOT retrieval systems while enabling simulated evaluation independent of CQA constraints.

While TOT retrieval builds on known-item retrieval research, our dataset and methodology provide new tools for evaluating and training retrieval systems to handle TOT queries more effectively. We anticipate its long-term value and plan to incrementally expand domain coverage through future TREC tracks, ensuring broader applicability and comprehensive evaluation in TOT retrieval research.

\textbf{Methodological Implications}.
While our findings show that both LLM- and human-elicited queries are effective, they serve complementary roles: LLMs offer scalability and efficiency, whereas human queries may provide authentic linguistic patterns and user behaviors. A hybrid approach can balance dataset diversity and efficiency, leading to more comprehensive evaluations of TOT retrieval systems.

Beyond TOT retrieval, our elicitation methods could support vague or exploratory search scenarios, where users struggle to articulate precise queries. Additionally, LLM-based query generation could aid low-resource domains, simulating real-world search behaviors where query logs are scarce or unavailable, broadening its impact across information retrieval research.

\textbf{Limitation and Future Work}.
A limitation of our current LLM-elicitation method is that prompts are domain-specific, limiting their generalizability. Future work should develop generalized prompting strategies to elicit TOT queries across diverse search contexts without extensive manual tuning.
Additionally, expanding the methodology to multi-turn or multi-modal interactions \cite{ch2025browsing} could better reflect real-world TOT search behavior, where users iteratively refine queries as they recover missing information. 

\section{Conclusion}\label{sec:conclusion}
This work introduces a novel approach to TOT query elicitation, leveraging LLMs and human participants to move beyond the limitations of CQA-based datasets. Through system rank correlation and linguistic similarity validation, we demonstrate that LLM- and human-elicited queries can effectively support the simulated evaluation of TOT retrieval systems. Our findings highlight the potential for expanding TOT retrieval research into underrepresented domains while ensuring scalability and reproducibility. The released datasets and source code provide a foundation for future research, enabling further advancements in TOT retrieval evaluation and system development.

\begin{acks}
We thank Ian Soboroff for implementing the in-house human query collection interface based on the MTurk pilot, and for coordinating contractor hiring and oversight at NIST.
We also appreciate Alfredo Gomez for helping collect validation queries for the Landmark and Person domains, and thank Alfredo Gomez, Athiya Deviyani, Jessica Huynh, and Shaily Bhatt for their feedback during pilot testing.
\end{acks}

\bibliographystyle{ACM-Reference-Format}
\balance
\bibliography{XX-references.bib}

\newpage
\appendix
\onecolumn  % Make the paper one column until it sees \twocolumn

\section{Wikipedia Summarization Prompts}
System prompt:
\begin{tcolorbox}
You are a text summarization assistant.
\end{tcolorbox}
\noindent
Domain-specific user prompts were used for the Movie, Landmark and Person domains.

\subsection{Movie Domain}
\begin{tcolorbox}
Please summarize the following description about a movie into two paragraphs:\\ \{Wikipedia Paragraphs\}.\\Please focus on the plots, and ignore the director and actor names.
\end{tcolorbox}

\subsection{Landmark Domain}
\begin{tcolorbox}
Please summarize the following description about a place into two paragraphs:\\ \{Wikipedia Paragraphs\}.
\end{tcolorbox}

\subsection{Person Domain}
\begin{tcolorbox}
Please summarize the following description about a person into two paragraphs:\\ \{Wikipedia Paragraphs\}.
\end{tcolorbox}

\newpage

\section{LLM-Elicitation Prompts (Version 6)}

\subsection{LLM-Elicitation Prompt for the Movie Domain}
\begin{tcolorbox}

Let's do a role play. You are now a person who watched a movie \{ToTObject\} a long time ago and forgot the movie's name. You are trying to recall the name by posting a verbose post in an online forum like Reddit describing the movie. Generate a post of length of about 200 words about the movie {ToTObject}. Your post must describe a vague memory of a movie without mentioning its exact name. People in the forum must have a hard time figuring out which movie you are looking for. The answer should be hard to find in search engines, so do not write too obvious search terms. I will provide you a basic information about the movie, and you have to follow the guidelines to generate a post.\\

Information about \{ToTObject\}:\\
\{WikipediaSummary\}\\

Guidelines:\\
MUST FOLLOW:\\
1. Reflect the imperfect nature of memory with phrases that express doubt or mixed recollections, avoiding direct phrases like "I'm not sure if it is true, but".\\
2. Do not specify any movie or actor names directly.\\
3. Refer to characters in a non-specific way using descriptions or roles rather than names.\\
4. Maintain a casual and conversational tone throughout the post, ensuring it sounds natural and engaging without using formal structures.\\
5. Provide vivid but ambiguous details to stir the reader's imagination while leaving them guessing.\\
6. Use the provided examples only as inspiration to craft a unique and engaging narrative, avoiding any direct replication of sample phrases.\\
7. Avoid using formal greetings such as "Hello" or "Hey everyone," and start directly with your post.\\

COULD FOLLOW:\\
1. Share a personal anecdote related to when or with whom you watched the movie, but avoid common phrases like "When I was young". Instead, think of unique ways to set the scene.\\
2. Focus on sensory details such as the overall mood, sounds, or emotional impact of the movie, using vivid descriptions.\\
3. Draw comparisons with other movies or familiar experiences but in a nuanced manner that doesn't directly echo well-known titles.\\
4. Introduce a few incorrect or mixed-up details to make the recollection seem more realistic and challenging to pinpoint.\\
5. Describe particular scenes or moments using ambiguous terms or partial descriptions.\\
6. Mention vaguely when and where you watched the movie, and encourage using less typical references than "10 years ago on TV".\\
7. Encourage responses with questions or prompts for help that sound genuine and open-ended.\\

Generate a post based on these guidelines.
\end{tcolorbox}

\newpage
\subsection{LLM-Elicitation Prompt for the Landmark Domain}
\begin{tcolorbox}
Let's do a role play. You are now a person who vaguely remembers a place called \{ToTObject\}. You are trying to recall the name of the place by posting a verbose post in an online forum like Reddit describing the place. Generate a post of around 200 words about the place \{ToTObject\}. Your post must describe a vague memory of the place without revealing its exact name. People on the forum must have a hard time figuring out which place you are looking for. The answer should be difficult to find in search engines, so avoid using obvious keywords. I will provide you with some basic information about the place, and you must follow the guidelines to create a post.\\

Information about \{ToTObject\}:\\
\{WikipediaSummary\}\\

Guidelines:\\
MUST FOLLOW:\\
1. Reflect the imperfect nature of memory with phrases that express doubt or mixed recollections, avoiding direct phrases like "I'm not sure if it is true, but".\\
2. Do not directly specify the name of the place.\\
3. Refer to the places in an ambiguous way using descriptions instead of names.\\
4. Maintain a casual and conversational tone throughout the post, making sure it sounds natural and engaging without using formal structures.\\
5. Provide vivid but ambiguous details to stir the reader's imagination while keeping them guessing.\\
6. Use the provided information only as inspiration to craft a unique and engaging narrative, avoiding any direct replication of the given phrases.\\
7. Start directly with your post, avoiding formal greetings like "Hello" or "Hey everyone."\\
8. Start directly with your post, without describing your state of mind like "So, there's this", "I remember", "I've been thinking about".\\

COULD FOLLOW:\\
1. Share a personal anecdote about your time at the place and the people you were with, but avoid common phrases like "When I was young." Instead, find unique ways to set the scene.\\
2. Focus on sensory details like the overall mood, sounds, and emotional impact of being in the place, using vivid descriptions.\\
3. Draw comparisons with other places or familiar experiences in a nuanced way that doesn't directly echo well-known locations.\\
4. Introduce a few incorrect or mixed-up details to make the recollection seem more realistic and harder to pinpoint.\\
5. Describe particular scenes or moments using ambiguous terms or partial descriptions.\\
6. End the post by encouraging responses with genuine, open-ended questions for help.\\

Generate a post based on these guidelines.

\end{tcolorbox}

\newpage
\subsection{LLM-Elicitation Prompt for the Person Domain}
\begin{tcolorbox}
Let's do a role play. You are now a person who vaguely remembers a public figure called \{ToTObject\}, but forgot the person's name. You are trying to recall the name by posting a verbose post in an online forum like Reddit describing the person. Generate a post of around 200 words about the person \{ToTObject\}. Your post must describe a vague memory of the person without revealing its exact name. People on the forum must have a hard time figuring out which person you are looking for. The answer should be difficult to find in search engines, so avoid using obvious keywords. I will provide you with some basic information about the person, and you must follow the guidelines to create a post.\\

Information about \{ToTObject\}:\\
\{WikipediaSummary\}\\

Guidelines:\\
MUST FOLLOW:\\
1. Reflect the imperfect nature of memory with phrases that express doubt or mixed recollections, avoiding direct phrases like "I'm not sure if it is true, but".\\
2. Do not directly specify the name of the person.\\
3. Refer to the person in an ambiguous way using descriptions instead of names.\\
4. Maintain a casual and conversational tone throughout the post, making sure it sounds natural and engaging without using formal structures.\\
5. Provide vivid but ambiguous details to stir the reader's imagination while keeping them guessing.\\
6. Use the provided information only as inspiration to craft a unique and engaging narrative, avoiding any direct replication of the given phrases.\\
7. Start directly with your post, avoiding formal greetings like "Hello" or "Hey everyone."\\
8. Start directly with your post, without describing your state of mind like "So, there's this", "I remember", "I've been thinking about".\\

COULD FOLLOW:\\
1. Share a personal anecdote related to the person, but avoid common phrases like "When I was young." Instead, find unique ways to set the scene.\\
2. Draw comparisons with other similar public figures in a nuanced way that doesn't directly echo well-known people.\\
3. Introduce a few incorrect or mixed-up details to make the recollection seem more realistic and harder to pinpoint.\\
4. Describe particular scenes or moments using ambiguous terms or partial descriptions.\\
5. End the post by encouraging responses with genuine, open-ended questions for help.\\

Generate a post based on these guidelines.

\end{tcolorbox}

\newpage

\section{Prompt for Linguistic Similarity Test in the Movie Domain}

System prompt:
\begin{tcolorbox}
You are an expert annotator.
\end{tcolorbox}

\noindent
User prompt influenced by \citet{arguello-movie-identification}:
\begin{tcolorbox}
Given a paragraph, you will annotate the paragraph with the following coding scheme. Read the paragraph and annotate each sentence following the coding scheme.
When outputting the response, follow the output JSON format.\\

Coding scheme:\\
movie: Sentence that describes characteristics of the movie itself (e.g., category, genre, tone, plot).
context: Context where the searcher previously engaged with the movie. References to when, where the searcher watched the movie, who they watched it with, and even world events that were happening around the time.\\
previous-search: Describes a previous attempt to find the movie title. References to failed search attempts, descriptions of search strategies that were unsuccessful.\\
social: Communicates a social nicety to the community of people (e.g., If you could help I’d really appreciate it).\\
uncertainty: Conveys uncertainty about information described (e.g., It was a foreign film I think either French or German, but I could be wrong).\\
opinion: Conveys an opinion or judgment about some aspect of the movie (e.g., Its pretty confusing all the way to the end when there’s only one surviving woman).\\
emotion: Conveys how the movie made the viewer feel (e.g., It was the first movie that kept me awake at night).\\
relative-comparison: Statement that draws one or more comparisons that require additional information in order to extract the meaning of the statement (e.g., the main character looks like Brad Pitt).\\

Paragraph:\\
\{Paragraph\}\\

Output JSON format:\\
\{\\
"<Sentence Number>": ["<code>"],\\
"<Sentence Number>": ["<code>", "<code>", ...],\\
"<Sentence Number>": []\\
...\\
\}\\

For each sentence number in a paragraph, you can put multiple appropriate code. However, if none of the codes match with a sentence, you can leave the list empty.
\end{tcolorbox}

\end{document}